\documentclass[prd,nofootinbib,preprint,superscriptaddress,twocolumn,10pt]{revtex4-1}
\pdfoutput=1
\usepackage{amsmath,amssymb}
\usepackage{epsfig}
\usepackage{graphicx}
\usepackage[usenames,dvipsnames]{color}
\usepackage{subfigure}
\usepackage{slashed}
\usepackage{comment}
\usepackage[normalem]{ulem}
\usepackage[colorlinks,citecolor=blue]{hyperref}
\usepackage{color}

\usepackage{multirow}

\usepackage{xcolor}
\usepackage{soul}

\begin{document}
%%%%%%%%%%%%%%%%%%%%%%%%%%%%%%%%%%%%%%%%
\title{Probing Left-Right Symmetry via Gravitational Waves from Domain Walls}
%%%%%%%%%%%%%%%%%%%%%%%%%%%%%%%%%%%%%%%%
\author{Debasish Borah}
\email{dborah@iitg.ac.in}
\affiliation{Department of Physics, Indian Institute of Technology Guwahati, Assam 781039, India}

\author{Arnab Dasgupta}
\email{arnabdasgupta@pitt.edu}
\affiliation{Pittsburgh Particle Physics, Astrophysics, and Cosmology Center, Department of Physics and Astronomy, University of Pittsburgh, Pittsburgh, PA 15206, USA}

%%%%%%%%%%%%%%%%%%%%%%%%%%%%%%%%%%%%%%%%%%%%%%%%%
\begin{abstract}
We study the possibility of probing the scale of left-right symmetry breaking in the context of left-right symmetric models (LRSM). In LRSM, the right handed fermions transform as doublets under a newly introduced $SU(2)_R$ gauge symmetry. This, along with a discrete parity symmetry $\mathcal{P}$ ensuring identical gauge couplings of left and right sectors make the model left-right symmetric, providing a dynamical origin of parity violation in electroweak interactions via spontaneous symmetry breaking. The spontaneous breaking of $\mathcal{P}$ leads to the formation of domain walls in the early universe. These walls, if made unstable by introducing an explicit parity breaking term, generate gravitational waves (GW) with a spectrum characterized by the wall tension or the spontaneous $\mathcal{P}$ breaking scale, and the explicit $\mathcal{P}$ breaking term. Considering explicit $\mathcal{P}$ breaking terms to originate from Planck suppressed operators provides one-to-one correspondence between the scale of left-right symmetry and sensitivities of near future GW experiments. This is not only complementary to collider and low energy probes of TeV scale LRSM but also to GW generated from first order phase transition in LRSM with different spectral shape, peak frequencies as well as symmetry breaking scales.
\end{abstract}
%%%%%%%%%%%%%%%%%%%%%%%%%%%%%%%%%%%%%%%%%%%%%%%%%%%%%%%%

\maketitle
%%%%%%%%%%%%%%%%%%%%%%%%%%%%%%%%%%%%%%%%%%%%%%%%%%%%%%%
\noindent
\textbf{\textit{Introduction:}} Left-right symmetric models (LRSM) \cite{Pati:1974yy, Mohapatra:1974hk, Mohapatra:1974gc, Senjanovic:1975rk, Senjanovic:1978ev, Mohapatra:1977mj, Mohapatra:1980qe, Mohapatra:1980yp, Lim:1981kv, Gunion:1989in,Deshpande:1990ip,Duka:1999uc, FileviezPerez:2008sr} have been one of the most well-motivated beyond standard model (BSM) frameworks studied extensively in the last few decades. The standard model (SM) gauge symmetry is extended to $ \rm SU(3)_c\times SU(2)_L\times SU(2)_R\times U(1)_{B-L}$ with the right handed fermions transforming also transforming as doublets under $SU(2)_R$ gauge symmetry. While such a setup treats left and right handed fermions on equal footing unlike in the SM, inclusion of right handed neutrinos automatically leads to possible ways of generating light neutrino masses, as confirmed by neutrino oscillation data \cite{Zyla:2020zbs}. In order to make the setup parity symmetric $SU(2)_L \leftrightarrow SU(2)_R$, an additional discrete $Z_2$ symmetry or left-right parity $\mathcal{P}$ is incorporated. It not only provides a dynamical origin of parity violation in weak interactions but can also be realised as an intermediate symmetry in popular grand unified theories (GUT) like SO(10). In order to break the LRSM gauge symmetry spontaneously into that of the SM, either a pair of scalar doublets \cite{Pati:1974yy, Mohapatra:1974hk, Mohapatra:1974gc, Senjanovic:1975rk, Senjanovic:1978ev, Bernard:2020cyi} or a pair of scalar triplets \cite{Mohapatra:1980qe, Mohapatra:1980yp, Lim:1981kv, Gunion:1989in,Deshpande:1990ip, Duka:1999uc, FileviezPerez:2008sr} are introduced. In both these scenarios, the scale of $SU(2)_R \times U(1)_{B-L}$ breaking coincides with the same $\mathcal{P}$ breaking\footnote{See \cite{Chang:1983fu, Chang:1984uy} for scenarios where these two symmetry breaking scales are decoupled.}. Apart from the details of fermion mass generations, these two classes of LRSM also have different experimental consequences at collider experiments like the large hadron collider (LHC) \cite{Maiezza:2010ic, ATLAS:2017jbq,ATLAS:2017eqx,CMS:2016gsl,CMS:2016ifc,CMS:2018hff, ATLAS:2019fgd, CMS:2019gwf}. The LRSM also can have interesting low energy probes some of which can be found in \cite{Tello:2010am, Li:2020flq, Dekens:2021bro, Borah:2016iqd, Borah:2015ufa, BhupalDev:2014qbx, Bambhaniya:2015ipg} and references therein.

In addition to collider and other low energy search prospects of TeV scale LRSM, we can also have interesting cosmological signatures in this model. For example, if light neutrinos are of Dirac type, then we can have additional relativistic degrees of freedom $\Delta N_{\rm eff}$ \cite{Borah:2020boy} which can be probed at future cosmic microwave background (CMB) experiments. It is also possible to have a strong first order phase transition in LRSM with observable consequences like stochastic gravitational waves (GW) \cite{Brdar:2019fur, Graf:2021xku, Barenboim:1998ib, Li:2020eun}. All these cosmological probes, like the laboratory ones, depend upon the scale of left-right symmetry breaking. In this work, we propose an alternative way to probe the scale of left-right symmetry via GW generated from collapsing domain walls (DW)\footnote{While this possibility was outlined in \cite{Craig:2020bnv} and \cite{Chun:2021brv} in the context of parity solution to strong CP problem and supersymmetric $SO(10)$ GUT respectively, here we discuss the details of different left-right models and resulting GW spectrum from collapsing DW.}. Such DW can arise due to spontaneous breaking of discrete parity symmetry $\mathcal{P}$ which is generic in LRSM. The walls can be made unstable by introducing an explicit $\mathcal{P}$ breaking term in the potential which causes a pressure difference across the DW, also known as the bias term $\Delta V$. Since such a bias term in the potential can arise in different ways, we first consider the scale of left-right symmetry breaking and bias term to be independent of each other and find the GW spectrum. We then consider a particular origin of bias term via Planck suppressed operators, motivated by quantum gravity arguments and show the sensitivity of near future experiments to different left-right symmetry breaking scales. Interestingly, we find our results not only complementary to collider and low energy experimental probes but also to GW from first order phase transition in LRSM discussed in earlier works. Contrary to GW from first order phase transition in LRSM with peak frequencies around $f \sim \mathcal{O}(10^{-2})-\mathcal{O}(1)$ Hz, GW from unstable domain walls can show up in the nano-Hz as well as intermediate frequency regime too depending upon the scale of symmetry breaking as well as the bias term. In addition, detection of GW from DW in future experiments can probe the scale left-right symmetry all the way from TeV to very high scale, in sharp contrast with GW from first order phase transition sensitive to the scale of symmetry breaking upto a few tens of TeV. It should also be noted that although the laboratory and other cosmological probes of LRSM depend upon the details of field content as well as the fermion mass generation, GW signature from collapsing DW is quite generic and depends on the scale of parity breaking only, if the bias terms arise from Planck suppressed operators.

\medskip
%\newpage
\noindent
\textbf{\textit{Left-right symmetric model:}}
The fermion and scalar contents of LRSM with scalar triplets \cite{Mohapatra:1980qe, Mohapatra:1980yp, Lim:1981kv, Gunion:1989in,Deshpande:1990ip, Duka:1999uc, FileviezPerez:2008sr} is given by
\begin{align}
{\rm Fermions:} \quad & Q_L \equiv (3,2,1,1/3),~~  Q_R \equiv (3,1,2,1/3), \nonumber \\
& \Psi_L \equiv (1,2,1,-1),~~  \Psi_R \equiv (1,1,2,-1) \nonumber \\
{\rm Scalars:} \quad &
\Phi \equiv (1,2,2,0), \nonumber \\
& \Delta_L \equiv (1,3,1,2), \quad \Delta_R \equiv (1,1,3,2)
\end{align}
where the numbers in the brackets are the quantum numbers corresponding to the LRSM gauge group $SU(3)_c \times SU(2)_L \times SU(2)_R \times U(1)_{B-L}$. The discrete left-right symmetry or parity is an additional $Z_2$ symmetry $\mathcal{P}$ under which the left and right sector fields get interchanged as
$$ Q_L \leftrightarrow Q_R, \Psi_L \leftrightarrow \Psi_R, \Delta_L \leftrightarrow \Delta_R, \Phi \leftrightarrow \Phi^{\dagger}. $$
This also ensures the equality of left and right sector gauge couplings $g_L = g_R$, in addition to relating the Yukawa and scalar potential couplings of these two sectors. As mentioned before, the symmetry group of LRSM can be realised as an intermediate stage symmetry when $SO(10)$ symmetry of GUT breaks down to the SM gauge symmetry. For example, $SO(10)$ can be broken down to $SU(3)_c \times SU(2)_L \times SU(2)_R \times U(1)_{B-L} \times \mathcal{P}$ by a scalar of {\bf 210} representation. For details of such $SO(10)$ breaking patterns with LRSM as intermediate symmetry, one may refer to \cite{Deshpande:1992au, Bertolini:2009es, Chakrabortty:2019fov, Held:2022hnw} and references therein.

The neutral component of the scalar triplet $\Delta_R$ acquires a non-zero vacuum expectation value (VEV) breaking both left-right symmetry $\mathcal{P}$ and $SU(2)_R \times U(1)_{B-L}$ gauge symmetry into $U(1)_Y$ of the SM. At a later stage, the electroweak gauge symmetry gets spontaneously broken to $U(1)_{\rm em}$ by the neutral components of scalar bidoublet. The symmetry breaking pattern is 
\begin{align}
 SU(2)_L \times SU(2)_R \times U(1)_{B-L} \times \mathcal{P} \quad \underrightarrow{\langle
\Delta_R \rangle} \nonumber \\
\quad SU(2)_L\times U(1)_Y \quad \underrightarrow{\langle \Phi \rangle} \quad U(1)_{\rm em}
\end{align}
If the scalar triplets are replaced by a pair of scalar doublets $H_{L,R}$, the neutral component of $H_R$ breaks $SU(2)_R \times U(1)_{B-L} \times \mathcal{P}$ into $U(1)_Y$ of the SM. 

In the model with scalar doublets, the scale of left-right breaking $M_R \equiv v_R$ can be related to $SU(2)_R$ charged gauge boson mass as
\begin{equation}
M^2_{W_R} \approx \frac{1}{4} g_R^2 v_R^2.
\end{equation}
In the scalar triplet version of LRSM, it is given by 
\begin{equation}
M^2_{W_R} \approx \frac{1}{2} g_R^2 v_R^2.
\end{equation}
Depending upon the scalar content of LRSM, the neutral heavy gauge boson $Z_R$ mass can be derived from $W_R$ mass. The direct search constraints from the LHC rules out $W_R$ mass upto a few TeV $M_{W_R} \geq 4-5$ TeV \cite{ATLAS:2017jbq,ATLAS:2017eqx,CMS:2016gsl,CMS:2016ifc,CMS:2018hff, ATLAS:2019fgd, CMS:2019gwf}.

\medskip
\noindent
\textbf{\textit{Domain walls and gravitational waves:}}
Topological defects like domain walls can form in the early universe when a discrete symmetry is broken spontaneously~\cite{Zeldovich:1974uw, Kibble:1976sj, Vilenkin:1981zs, Kibble:1982dd, Lazarides:1981fv, Vilenkin:1984ib}. As the energy density of DW falls with the expansion of the universe at a slower rate compared to that of ordinary radiation or matter, they can start dominating the energy density of the universe and can ruin the successful predictions of standard cosmology \cite{Press:1989yh}. However, this can be prevented if the DW are made unstable or diluted or if the probability distribution for initial field fluctuations is asymmetric \cite{Coulson:1995nv, Krajewski:2021jje}. Let us consider the example of a $Z_2$-odd scalar singlet $\phi$, having the potential
\begin{align}
	V(\phi) = \frac{\lambda_\phi}{4}(\phi^2-u^2)^2\,,
	\label{V}
\end{align}
having two different vacua $\langle \phi \rangle = \pm u$. One can find a static solution of the equation of motion after imposing a boundary condition such that the two vacua are realized at $x \to \pm \infty$,
\begin{align}
	\phi({\bf x}) = u \tanh\left( \sqrt{\frac{\lambda_\phi}{2}} u x \right),
\end{align}
which represents a DW extended along the $x = 0$ plane. The DW width $\delta$ is approximately the inverse of the mass of $\phi$ at the potential minimum that is, $\delta \sim m_\phi^{-1} = (\sqrt{2\lambda_\phi} u)^{-1}$. Another key parameter, known as the DW tension is given by
\begin{align}
	\sigma = \int_{-\infty}^{\infty} dx \,\rho_\phi = \frac{2\sqrt 2}{3}\sqrt{\lambda_\phi} u^3 = \frac{2}{3}m_\phi u^2\,,
\end{align}
where $\rho_\phi = \frac{1}{2} |\nabla \phi|^2 + V(\phi)$ is the (static) energy density of $\phi$. For $m_\phi \sim u$, the tension of the wall can be approximated as $\sigma \sim u^3$.

\begin{figure*}
\includegraphics[scale=0.6]{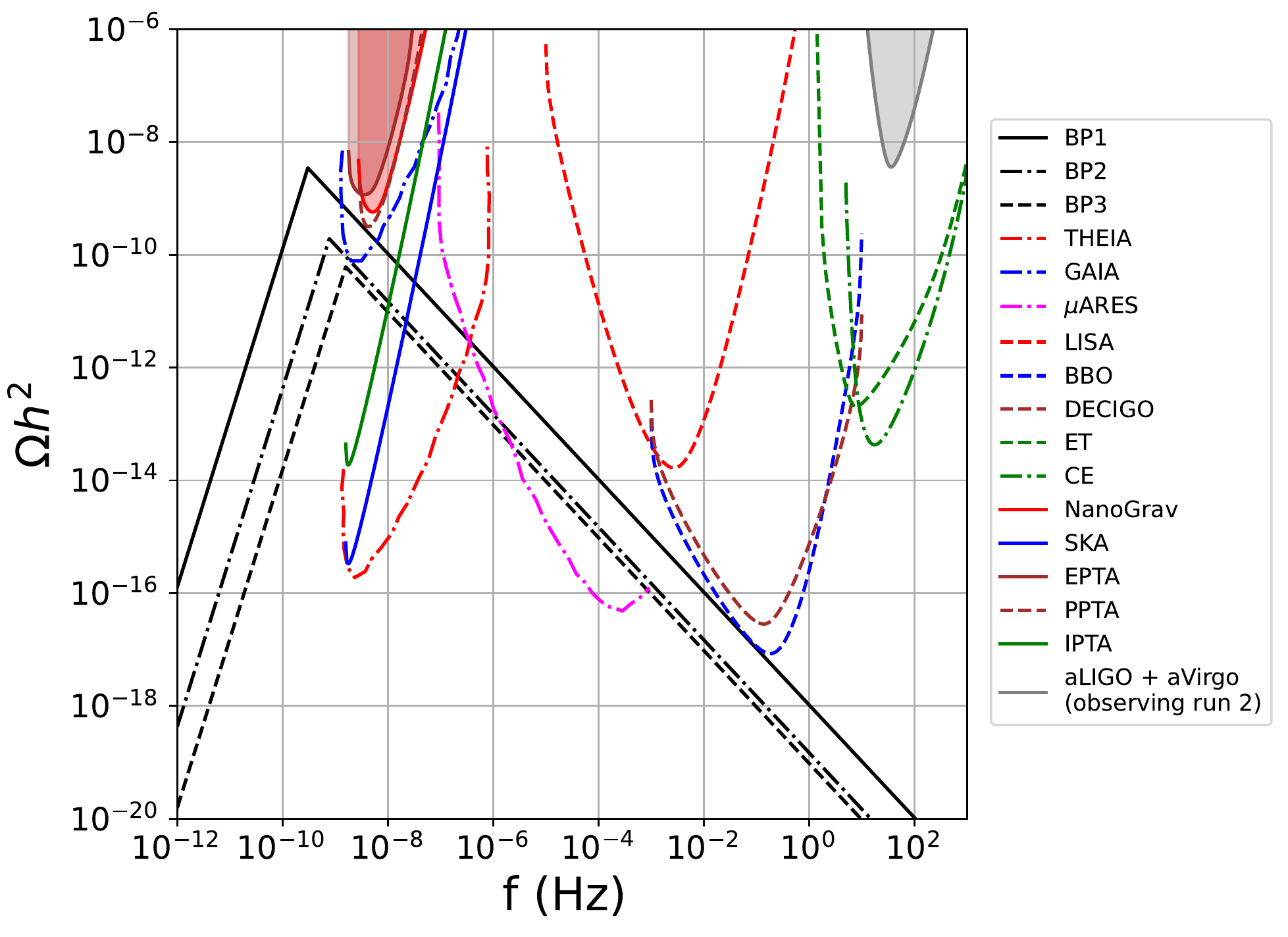}
\caption{Gravitational wave spectrum from collapsing domain walls in LRSM for three different benchmark combination of $(M_{W_R}, \Delta V)$ given in table \ref{tab:1}. Different coloured curves show the sensitivities from GW search experiments like LISA, BBO, DECIGO, HL (aLIGO), ET, CE, NANOGrav, SKA, EPTA, PPTA, IPTA, GAIA, THEIA and $\mu$ARES.}
\label{fig:1} 
\end{figure*} 

Since we consider the walls to form in a radiation dominated universe after inflation, we need to make them unstable so that they disappear eventually. As pointed out long ago \cite{Zeldovich:1974uw, Vilenkin:1981zs, Sikivie:1982qv, Gelmini:1988sf}, such DW can be made unstable simply by introducing a pressure difference across the walls and such pressure difference can originate from a small explicit symmetry breaking term in the potential. Such a pressure difference or bias term $\Delta V$ should be large enough to ensure the disappearance of the walls before the big bang nucleosynthesis (BBN) epoch that is, $t_{\rm BBN} > t_{\rm dec} \approx \sigma/\Delta V$. We also ensure that the walls disappear before dominating the universe $t_{\rm dec} < t_{\rm dom}$, where $t_{\rm dom} \sim  M^2_{\rm Pl}/ \sigma$ is the typical epoch of domain wall domination. Both of these criteria put a lower bound on the bias term $\Delta V$. However, the bias term $\Delta V$ can not be arbitrarily large due to the requirement of percolation of both the vacua (separated by DW) whose relative population can be estimated as $p_+/p_- \simeq e^{-4\Delta V/(\lambda_\phi u^4)}$ \cite{Gelmini:1988sf}. We choose $\Delta V$ such that the required percolation can be achieved trivially \cite{Gelmini:1988sf}. Such unstable DW can annihilate and radiate their energy via stochastic gravitational waves, the details of which has been studied in several works~\cite{Kadota:2015dza, Hiramatsu:2013qaa, Krajewski:2016vbr, Nakayama:2016gxi, Dunsky:2020dhn, Babichev:2021uvl, Ferreira:2022zzo, Deng:2020dnf, Gelmini:2020bqg, Saikawa:2017hiv}. The amplitude of such GW at peak frequency $f_{\rm peak}$ can be estimated as~\cite{Kadota:2015dza, Hiramatsu:2013qaa}
\begin{align}
    \Omega_{\rm GW}h^2 (t_0) \rvert_{\rm peak} & \simeq 5.2 \times 10^{-20} \tilde{\epsilon}_{\rm gw} A^4 \left ( \frac{10.75}{g_*} \right)^{1/3} \nonumber \\
    & \times \left ( \frac{\sigma}{1 \, {\rm TeV}^3 } \right)^4  \left ( \frac{1 \, {\rm MeV}^4}{\Delta V} \right)^2\,,
\end{align}
with $t_0$ being the present time and $g_*$ is the relativistic degrees of freedom at the epoch of GW emission which is assumed to be same as the epoch of DW collapse $t_{\rm dec} \sim \sigma/\Delta V$. Away from the peak, the amplitude varies as 
\begin{align}
	\Omega_{\rm GW} \simeq \Omega_{\rm GW}\rvert_{\rm peak} \times 
	\begin{cases}
		\displaystyle{\left( \frac{f_{\rm peak}}{f} \right)} & {\rm for}~~f>f_{\rm peak}\\
		&\\
		\displaystyle{\left( \frac{f}{f_{\rm peak}} \right)^3 }& {\rm for}~~f<f_{\rm peak}
	\end{cases}\,,
\end{align}
where the peak frequency is given by
\begin{align}
    f_{\rm peak} (t_0) & \simeq 3.99 \times 10^{-9} \, {\rm Hz} A^{-1/2} \nonumber \\
    & \times \left ( \frac{ 1\, {\rm TeV}^3}{\sigma} \right)^{1/2} \left ( \frac{\Delta V}{1\, {\rm MeV}^4} \right)^{1/2}\,.
\end{align}
In the above expressions, $A$ is the area parameter~\cite{Caprini:2017vnn, Paul:2020wbz} $\simeq 0.8$ for DW arising from $Z_2$ symmetry breaking like we have in LRSM, and $\tilde{\epsilon}_{\rm gw}$ is the efficiency parameter $\simeq$ 0.7~\cite{Hiramatsu:2013qaa}. Since the GW amplitude at peak frequency increases with DW tension or equivalently, the $Z_2$ symmetry breaking scalar VEV, one can derive an upper bound on this VEV from the requirement of GW not to generate excess radiation or relativistic degrees of freedom. Cosmological observations from the PLANCK satellite and the corresponding cosmic microwave background (CMB) limits on additional effective relativistic degrees of freedom $\Delta N_{\rm eff}$ result in an upper bound $\Omega_{\rm GW} h^2 \lesssim 10^{-6}$~\cite{Boyle:2007zx,Stewart:2007fu,Pagano:2015hma, Lasky:2015lej, Aghanim:2018eyx}. Similar but slightly weaker bounds can be applied from the BBN limits on $\Delta N_{\rm eff}$ as well. It is worth mentioning that in the above description, we have ignored the friction effects which can be present between the walls and the background thermal plasma \cite{Nakayama:2016gxi, Galtsov:2017udh}. Such friction effects can be significant if the field constituting the wall has large couplings with the SM bath particles like the SM Higgs, leading to a suppression in resulting GW amplitude compared to the friction-less scenario discussed above. We neglect such frictional effects assuming that the parity breaking scalar coupling with the SM bath to be tiny \cite{Babichev:2021uvl}. This is valid in the sense that after the formation of DW or left-right breaking, the only fields to which the parity breaking scalar couples significantly decouple from the plasma due to their heavy masses.

\begin{figure}
\includegraphics[scale=0.56]{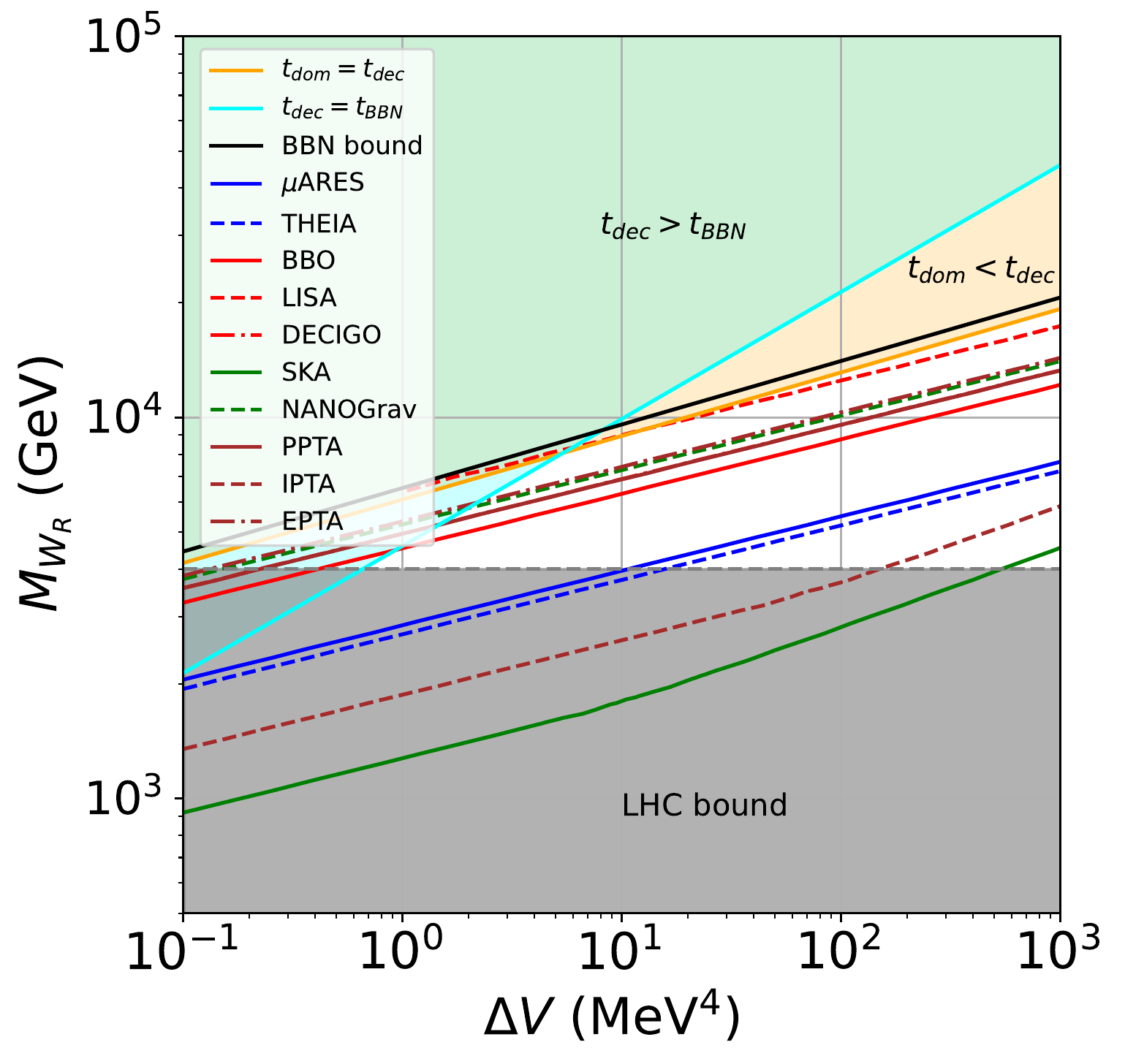}
\caption{Contours of SNR = 10 for different GW experiments shown in $M_{W_R}-\Delta V$ plane. The region above these contours correspond to SNR $>$ 10 for the corresponding experiment. The region above the black solid line corresponds to $\Omega_{\rm GW} h^2 \gtrsim 10^{-6}$ or equivalently $\Delta N_{\rm eff} >0.3$ and hence disfavoured. The shaded regions in upper part correspond to DW decaying after BBN $(t_{\rm dec} > t_{\rm BBN})$ and dominating before decaying $(t_{\rm dom} < t_{\rm dec})$ respectively and hence disfavoured. The grey shaded region at the bottom is disfavoured from the LHC bounds.}
\label{fig:2} 
\end{figure} 

\begin{figure*}
\includegraphics[scale=0.6]{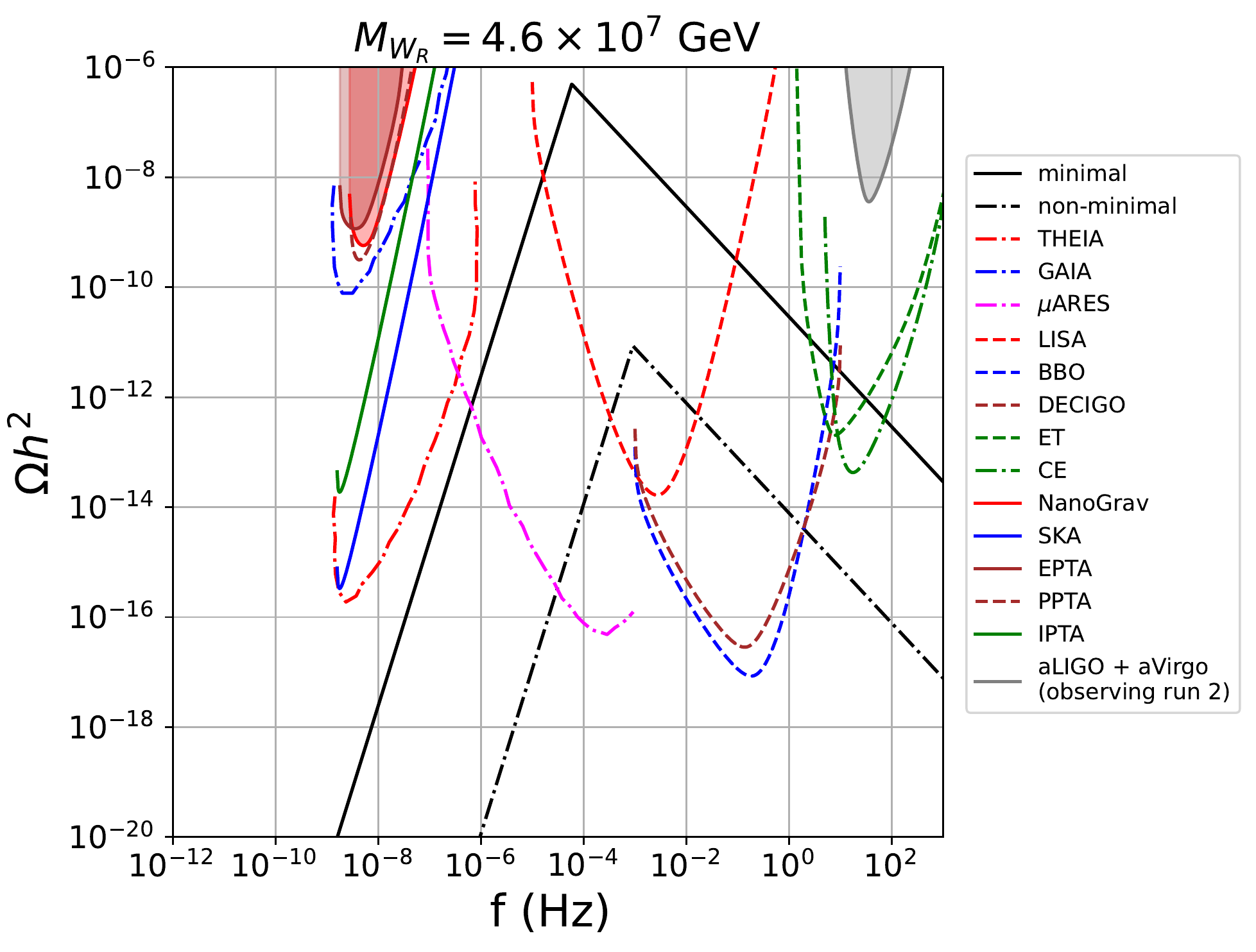}
\caption{Gravitational wave spectrum from collapsing domain walls in LRSM for a particular benchmark value of $M_{W_R}$ but with two different ways of generating bias terms, labelled as minimal and non-minimal respectively (see text for details). Different coloured curves show the sensitivities from GW search experiments like LISA, BBO, DECIGO, HL (aLIGO), ET, CE, NANOGrav, SKA, EPTA, PPTA, IPTA, GAIA, THEIA and $\mu$ARES.}
\label{fig:2a} 
\end{figure*} 

\begin{figure}
\includegraphics[scale=0.4]{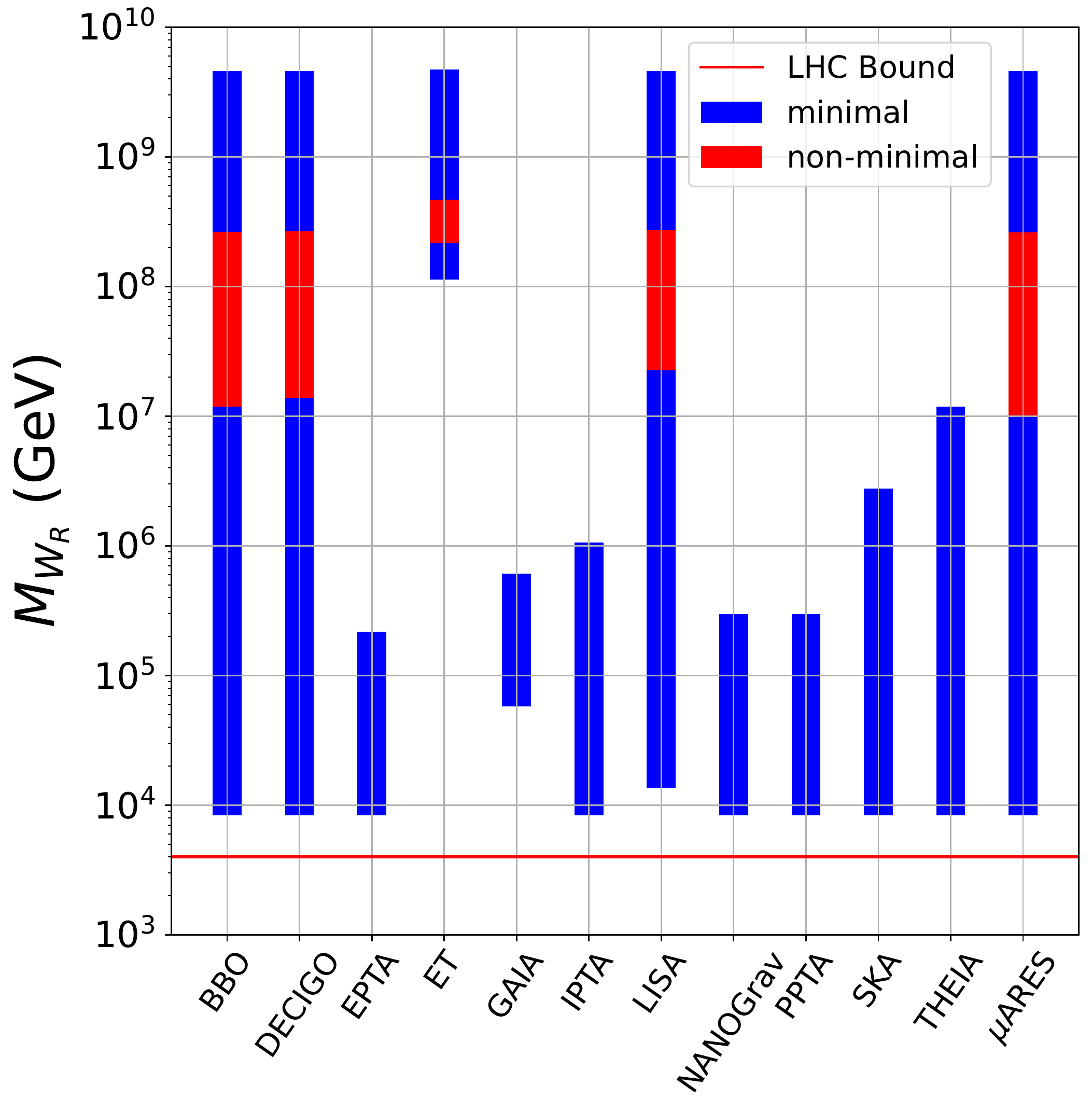}
\caption{Sensitivities of different GW experiments to the corresponding scale of left-right symmetry breaking or $M_{W_R}$ with SNR $\geq$ 10 by considering the bias term to originate from Planck suppressed dimension six and dimension five operators, labelled as minimal and non-minimal model respectively. The region below the red solid line is disfavoured from the LHC bounds. The bounds on parameter space from $t_{\rm dec} < t_{\rm BBN}, t_{\rm dom} > t_{\rm dec}$ are also satisfied.}
\label{fig:3} 
\end{figure} 

\medskip
\noindent
\textbf{\textit{Domain walls in LRSM:}} In both the versions of LRSM discussed above, since the left-right symmetry or parity $\mathcal{P}$ is spontaneously broken, it leads to the formation of domain walls. This has been studied in the context of LRSM as well as its $SO(10)$ embedding in several earlier works \cite{Mishra:2009mk, Borah:2011qq, Borah:2011aw, Borah:2012vk, Banerjee:2020zxw, Banerjee:2018hlp, Garg:2018trf, Chakrabortty:2019fov}. As pointed out earlier, a small explicit parity breaking term can make these DW unstable recovering the standard cosmology by the BBN epoch. In LRSM, this can be done by introducing higher dimensional gauge invariant but parity breaking operators. These operators can be suppressed by the Planck scale based on the argument that any generic theories of quantum gravity should not respect global symmetries: both discrete and continuous \cite{Abbott:1989jw, Kallosh:1995hi, Hawking:1975vcx}.  It was also pointed 
out in \cite{Rai:1992xw,Lew:1993yt} that Planck scale suppressed non-renormalisable operators can be a source of domain wall instability. Gauge structure of the underlying theory dictates the structure of these non-renormalisable operators. Role of such operators on neutrino mass and mixing has been discussed in \cite{Borah:2013mqa, Ibarra:2018dib, Borah:2019ldn}.

In either of the versions of LRSM mentioned above (to be referred to as the minimal model hereafter), such operators can arise only at dimension six level, given by
\begin{equation}
V_{\text{NR}} \supset f_L \frac{(\Sigma^{\dagger}_L \Sigma_L )^3}{M^2_{\rm Pl}} + f_R \frac{(\Sigma^{\dagger}_R \Sigma_R )^3}{M^2_{\rm Pl}}
\end{equation}
where $\Sigma_{L,R} \equiv \Delta_{L,R}, H_{L,R}$ depending upon the type of LRSM. If the vacuum with right sector fields getting VEV $\sim M_R$,  the vacuum energy corresponding to the Planck suppressed term is 
\begin{equation}
\rho^R_{\rm eff} \sim \frac{f_R}{M^2_{\rm Pl}}M^6_R.
\end{equation}
Similarly, if left-sector fields acquire non-zero VEV we get
\begin{equation}
 \rho^L_{\rm eff} \sim \frac{f_L}{M^2_{\rm Pl}}M^6_L.
 \end{equation}
The left-right symmetry or parity makes it equally likely for left and right sector fields to acquire the same VEV and hence $M_L = M_R$. Therefore, the effective energy difference across the walls separating these two vacua is given by
\begin{equation}
\delta \rho =\Delta V\sim \frac{(f_L-f_R)}{M^2_{\rm Pl}}M^6_R.
\end{equation}
Since only the Planck suppressed terms break parity explicitly, there arises no energy difference due to renormalisable terms of the scalar potential. For order one coefficients $f_{L,R} \sim \mathcal{O}(1)$, the bias term is $\Delta V \sim M^6_R/M^2_{\rm Pl}$ in both the versions of LRSM discussed above.

For non-minimal scalar content, it is possible to realise such explicit $\mathcal{P}$ breaking operators at dimension five level too \cite{Borah:2012vk}. For example, introducing a pair of real scalar triplets $\Omega_L \equiv (1, 3, 1, 0), \Omega_R \equiv (1, 1, 3, 0) $ in either of the minimal model can lead to a two step breaking of LRSM into that of the SM gauge symmetry. For example, in such a non-minimal LRSM with scalar doublets $H_{L,R}$ and real scalar triplets $\Omega_{L,R}$, the symmetry breaking chain will be as follows
\begin{align}
 SU(2)_L \times SU(2)_R \times U(1)_{B-L} \times \mathcal{P} \quad \underrightarrow{\langle
\Omega_R \rangle} \nonumber \\
\quad SU(2)_L\times U(1)_R \times U(1)_{B-L} \quad \underrightarrow{\langle H_R \rangle} \nonumber \\
\quad SU(2)_L \times U(1)_Y \quad \underrightarrow{\langle \Phi \rangle} \quad U(1)_{\rm em}.
\end{align}
In such a non-minimal model, one can have explicit parity breaking Planck suppressed operators at dimension five as well as follows
\begin{equation}
V_{\text{NR}} \supset f_L \frac{{\rm Tr}[(\Omega^2_L] (H^\dagger_L \Omega_L H_L)}{M_{\rm Pl}} + f_R \frac{{\rm Tr}[(\Omega^2_R] (H^\dagger_R \Omega_R H_R)}{M_{\rm Pl}}.
\end{equation}
Due to lower dimension compared to the minimal model, we can have a larger bias term in such non-minimal models.

\begin{table}
    \centering
%    \footnotesize{
    \begin{tabular}{|c|c|c|}
    \hline
     & $M_{W_R}$ (TeV)  & $\Delta V$ $({\rm MeV}^4)$ \\
    \hline  
    BP1 & $5.98$  & $10$  \\
    BP2 & $6.90$   & $10^2$ \\
    BP3 & $9.20$   & $10^3$ \\
    \hline
    \end{tabular}
%    \label{tab:Lepto}
    \caption{Details of the benchmark parameters used to generate the GW spectrum from domain walls in Fig. \ref{fig:1}.}
  \label{tab:1}
\end{table}

Since the bias term can originate in different ways, we first consider $M_{W_R}$ and $\Delta V$ to be independent of each other and show the GW spectrum for three different benchmark combinations in Fig. \ref{fig:1}. The details of the benchmark parameters are given in table \ref{tab:1}. While these choices of $M_{W_R}$ and $\Delta V$ in table \ref{tab:1} are arbitrary, they show the dependence of GW spectrum and peak frequencies on these two key parameters. The experimental sensitivities of NANOGrav \cite{McLaughlin:2013ira}, SKA \cite{Weltman:2018zrl}, GAIA \cite{Garcia-Bellido:2021zgu}, THEIA \cite{Garcia-Bellido:2021zgu}, $\mu$ARES \cite{Sesana:2019vho}, LISA\,\cite{AmaroSeoane2012LaserIS}, DECIGO \cite{Kawamura:2006up}, BBO\,\cite{Yagi:2011wg}, ET\,\cite{Punturo_2010}, CE\,\cite{LIGOScientific:2016wof} and aLIGO \cite{LIGOScientific:2014pky}, PPTA \cite{Manchester_2013}, IPTA \cite{Hobbs_2010},  EPTA \cite{Kramer_2013} are shown as curves of different styles. Although the spectrum has peak type features, the peak frequencies remain near the nano-Hz regime (within the ballpark of pulsar timing array (PTA) based experiments) in sharp contrast with GW originating from first order phase transitions in LRSM where peak frequencies remain on the higher side near the ballpark of experiments like LISA \cite{Brdar:2019fur, Graf:2021xku, Barenboim:1998ib, Li:2020eun}. In fact, for typical choices of model parameters, the peak frequencies remain outside the reach of all the experiments with one of the arms of the GW spectrum falling within sensitivity curves of several experiments from nano-Hz to Hz regime.

In Fig. \ref{fig:2} we show the contours in $M_{W_R}-\Delta V$ plane for different GW experiment corresponding to the signal-to-noise ratio (SNR) at respective GW experiments to be more than 10. The SNR is defined as~\cite{Dunsky:2021tih, Schmitz:2020syl} 
\begin{equation}
\rho = \sqrt{\tau\,\int_{f_\text{min}}^{f_\text{max}}\,df\,\left[\frac{\Omega_\text{GW}(f)\,h^2}{\Omega_\text{expt}(f)\,h^2}\right]^2}\,, 
\end{equation}
with $\tau$ being the observation time in years for a particular GW detector. In each of these contours, the SNR for the respective GW experiment correspond to 10 assuming that the experiment will operate for at least four years. The region above each of these contours corresponds to SNR $>$ 10 for the respective GW experiment while the region above the solid black line is ruled out by BBN as well as CMB limits on $\Delta N_{\rm eff}$. While we show the parameter space for GW experiments BBO \cite{Yagi:2011wg}, LISA\,\cite{AmaroSeoane2012LaserIS}, DECIGO \cite{Kawamura:2006up}, PPTA \cite{Manchester_2013}, IPTA \cite{Hobbs_2010},  EPTA \cite{Kramer_2013}, SKA \cite{Weltman:2018zrl}, THEIA \cite{Garcia-Bellido:2021zgu}, $\mu$ARES \cite{Sesana:2019vho} and NANOGrav \cite{McLaughlin:2013ira} only, for remaining experiments like ET, CE, GAIA the required SNR can not be obtained in the chosen range of $M_{W_R}-\Delta V$ shown in Fig. \ref{fig:2}. We also shade the disfavored regions in upper part of Fig. \ref{fig:2} from the requirements that the domain walls decay before BBN epoch as well as before dominating the energy density of the universe. 

We then consider two possible origins of the bias term from dimension six and dimension five Planck suppressed operators respectively, as discussed above and evaluate the range of left-right symmetry breaking scale which can be probed by future GW experiments with SNR $>$ 10. In Fig. \ref{fig:2a}, we first show the GW spectrum to compare the minimal and non-minimal model where the bias terms originate from dimension six and dimension five Planck suppressed operators respectively. For the non-minimal model, we consider the second step of symmetry breaking to occur at 10 TeV while $SU(2)_R \times \mathcal{P}$ breaks at a higher scale which generates $W_R$ mass. Since the non-minimal model has a larger bias term, the corresponding GW amplitude goes down as expected. In Fig. \ref{fig:3}, we show the range of $M_{W_R}$ or equivalently the $\mathcal{P}$ breaking scale within future experimental sensitivities of SNR $>$ 10, for both the minimal and non-minimal models. Clearly, one can probe from TeV to very high scale left-right symmetry by detecting GW from collapsing DW. In addition to the sensitivity of low frequency PTA based experiments to left-right symmetry breaking scale mentioned earlier, another sharp contrast from first order phase transition based GW probe of LRSM \cite{Brdar:2019fur, Graf:2021xku, Barenboim:1998ib, Li:2020eun} is the fact that one can probe very high scale left-right symmetry with GW from DW. Usually, GW originating from first order phase transition can be detected in future experiments to probe left-right symmetry upto around tens of TeV \cite{Brdar:2019fur, Graf:2021xku, Barenboim:1998ib, Li:2020eun}. 

\medskip
\noindent
\textbf{\textit{Conclusions:}} We have proposed a gravitational wave based probe of left-right symmetry breaking by future detection of stochastic GW background originating from collapsing domain walls formed as a result of discrete left-right symmetry or parity breaking. The domain walls are made unstable by introducing a small explicit parity breaking term into the potential, which can also arise from Planck suppressed higher dimensional operators motivated from quantum gravity arguments. Such explicit breaking introduces a bias term or equivalently a pressure difference across the walls eventually making them disappear. Considering the scale of left-right symmetry and the bias term to be independent of each other, we show the sensitivities of future GW experiments to the scale of left-right symmetry for a given bias term. We then consider the bias term to be dependent on the scale of left-right symmetry via higher dimensional operators and show that the future GW experiments can probe left-right symmetry braking all the way from TeV to very high scale. This not only offers a complementary probe to usual collider or low energy frontier experiments but also to GW from first order phase transition in LRSM. While future GW experiments operating only in the high frequency range around LISA sensitivity can probe left-right symmetry breaking upto a few tens of TeV, our proposal allows the probe of left-right symmetry scale in a much wider range and also in a wide range of GW experiments sensitive to frequencies as low as nano-Hz regime. It can also be distinguished from GW probes of other high scale scenarios with topological defect like cosmic strings being responsible for generating the stochastic GW background with contrasting spectral shapes \cite{Dror:2019syi, Blasi:2020wpy, Fornal:2020esl, Samanta:2020cdk, Borah:2022byb}. In LRSM itself, one can have cosmic string formation as a result of extended gauge symmetry breaking \cite{Yajnik:1998sw}. These strings can generate stochastic GW background with a characteristic spectrum which can be within the reach of near future GW detectors if the scale of symmetry breaking is sufficiently high \cite{Vilenkin:1981bx,Turok:1984cn}. Depending upon the symmetry breaking scales, both domain walls and cosmic strings can individually generate observable GW background predicting a spectrum which results from a combination of peak type feature of domain wall generated ones and the scale-invariant ones generated by cosmic strings. We leave such detailed studies on multiple sources of GW within LRSM to future works.

\acknowledgements
DB would like to acknowledge the hospitality at PITT-PACC, University of Pittsburgh where this work was completed.
\twocolumngrid
%\bibliographystyle{apsrev}
%\bibstyle{apsrev}
%\bibliography{ref.bib, ref_aks.bib, ref1.bib}

\end{document}